\documentclass[runningheads]{llncs}
\usepackage{graphicx}
\graphicspath{{figures/}}

\usepackage{color}
\usepackage{amsmath}       
\usepackage{breqn}
\usepackage{subcaption}
\usepackage[noadjust]{cite}
\usepackage{amssymb}

\begin{document}
\title{CorrSigNet: Learning CORRelated Prostate Cancer SIGnatures from Radiology and Pathology Images for Improved Computer Aided Diagnosis}
\titlerunning{CorrSigNet}
%
\author{Indrani Bhattacharya \inst{1} \and Arun Seetharaman \inst{2} \and Wei Shao \inst{1} \and Rewa Sood \inst{2} \and Christian A. Kunder \inst{4} \and Richard E. Fan \inst{3} \and Simon John Christoph Soerensen \inst{3,5} \and Jeffrey B. Wang \inst{1} \and Pejman Ghanouni\inst{1,3} \and Nikola C. Teslovich \inst{3} \and James D. Brooks \inst{3} \and Geoffrey A. Sonn \inst{1,3} \and Mirabela Rusu \inst{1} \thanks{We thank the Departments of Radiology and Urology at Stanford University, for their support for this work.}}
\authorrunning{I. Bhattacharya et al.}
\institute{Department of Radiology, School of Medicine, Stanford University, CA 94305, USA. \and
Department of Electrical Engineering, Stanford University, CA 94305, USA. \and
Department of Urology, School of Medicine, Stanford University, CA 94305, USA.
\and Department of Pathology, School of Medicine, Stanford University, CA 94305, USA. \and Department of Urology, Aarhus University, Aarhus, Denmark.}
\maketitle              
\begin{abstract}
Magnetic Resonance Imaging (MRI) is widely used for screening and staging prostate cancer. However, many prostate cancers have subtle features which are not easily identifiable on MRI, resulting in missed diagnoses and alarming variability in radiologist interpretation. Machine learning models have been developed in an effort to improve cancer identification, but current models localize cancer using MRI-derived features, while failing to consider the disease pathology characteristics observed on resected tissue. In this paper, we propose CorrSigNet, an automated two-step model that localizes prostate cancer on MRI by capturing the pathology features of cancer. First, the model learns MRI signatures of cancer that are correlated with corresponding histopathology features using Common Representation Learning. Second, the model uses the learned correlated MRI features to train a Convolutional Neural Network to localize prostate cancer. The histopathology images are used only in the first step to learn the correlated features. Once learned, these correlated features can be extracted from MRI of new patients (without histopathology or surgery) to localize cancer. We trained and validated our framework on a unique dataset of 75 patients with 806 slices who underwent MRI followed by prostatectomy surgery. We tested our method on an independent test set of 20 prostatectomy patients (139 slices, 24 cancerous lesions, 1.12M pixels) and achieved a per-pixel sensitivity of 0.81, specificity of 0.71, AUC of 0.86 and a per-lesion AUC of $0.96 \pm 0.07$, outperforming the current state-of-the-art accuracy in predicting prostate cancer using MRI. 

\end{abstract}
\section{Introduction}

Early localization of prostate cancer from MRI is crucial for successful diagnosis and local therapy. However, subtle differences between benign conditions and cancer on MRI often make human interpretation challenging, leading to missed diagnoses and an alarming variability in radiologist interpretation. Human interpretation of prostate MRI suffers from low inter-reader agreement (0.46-0.78)\cite{barentsz2016synopsis} and high variability in reported sensitivity (58-98\%) and specificity (23-87\%) \cite{ahmed2017diagnostic}.

Predictive models can help standardize radiologist interpretation, but current models \cite{viswanath2012central, sumathipala2018prostate,litjens2014computer, armato2018prostatex, viswanath2019comparing, cao2019joint} often learn from MRI only, without considering the disease pathology characteristics. These approaches derive MRI features that are agnostic to the biology of the tumor. Moreover, current predictive models mostly use inaccurate labels (either from biopsies \cite{armato2018prostatex} that suffer from sampling errors, or cognitive registration of pre-operative MRI with digital histopathology images of surgical specimens,  where a radiologist retrospectively outlines the lesions on MRI \cite{sumathipala2018prostate}). MRI under-estimates the tumor size \cite{priester2017magnetic}, making outlines on MRI alone insufficient to capture the entire extent of disease. Furthermore, it is challenging to outline the \textasciitilde 20\% of tumors that are not clearly seen on MRI, even when using histopathology images as reference \cite{barentsz2016synopsis}. These MRI-based models use a variety of techniques including traditional classifiers with hand-crafted and radiomic features \cite{viswanath2012central, litjens2014computer, viswanath2019comparing}, as well as deep learning based models \cite{sumathipala2018prostate, cao2019joint}. The current state-of-the-art approach \cite{sumathipala2018prostate} to predict a cancer probability map for the entire prostate uses a Holistically Nested Edge Detection (HED) \cite{HED} algorithm. 

In this paper, we propose CorrSigNet, a two-step approach for predicting prostate cancer using MRI. First, CorrSigNet leverages spatially aligned radiology and histopathology images of prostate surgery patients to learn MRI cancer signatures that correlate with features extracted from the histopathology images. Second, CorrSigNet uses these correlated MRI signatures to train a predictive model for localizing cancer when histopathology images are not available, e.g.~before surgery. This approach enables learning MRI signatures that capture tumor biology information from surgery patients with histopathology images, and then translating those learned signatures for prediction in patients without surgery/biopsy. Prior studies lack such correlation analysis of the two modalities. Our approach shows improved prostate cancer prediction compared to the current state-of-the-art method \cite{sumathipala2018prostate}. 

\section{Proposed Method}
\subsection{Dataset}\label{sec:preprocess}
We used 95 prostate surgery patients with pre-operative multi-parametric MRI (T2-weighted and Apparent Diffusion Coefficient) and post-operative digitized histopathology images. Custom 3D printed molds were used to ensure that excised prostate tissue was sectioned in the same plane as the T2-weighted (T2W) MRI. An expert pathologist annotated cancer on the histopathology images. We spatially aligned the pre-operative MRI and digitized histopathology images of the excised tissue via the state-of-the-art RAPSODI registration platform \cite{mirabelaregistration}. RAPSODI achieved a Dice similarity coefficient of 0.98$\pm$0.01 for the prostate, prostate boundary Hausdorff distance of
1.71$\pm$0.48 mm, and urethra deviation of 2.91$\pm$1.25 mm between registered histopathology images and MRI. Such careful registration of radiology and pathology images of the prostate enabled (1) correlation analysis of the two modalities at a pixel-level, and (2) accurate mapping of cancer labels from pathology to radiology images. We considered multiple slices per patient (average 7 slices/patient) irrespective of cancer size. Slices with missing cancer annotations were discarded during training. The dataset included some patients with cancer that had extra prostatic extensions, but our analysis was focused only on cancers inside the prostate.

\subsection{Data Pre-processing} 
We smoothed the histopathology images with a Gaussian filter with $\sigma = 0.25$ to prevent downsampling artifacts, padded and then downsampled them to $224 \times 224$, resulting in an X-Y resolution of $0.29 \times 0.29~mm^2$. 
We projected and resampled the T2W and ADC images, prostate masks, and cancer labels on the corresponding downsampled histopathology images, such that they also had the same X-Y resolution of $0.29 \times 0.29~mm^2$. This ensured that each pixel in each modality represented the same physical area.

Since MRI intensities vary significantly between scanners and scanning protocols, we standardized the T2W and ADC intensities using the histogram alignment approach proposed by Ny{\'u}l et al.~\cite{nyul2000new}. We used prostate masks to standardize intensities within the prostate, and then applied the learned transformation to the image region beyond the prostate. After intensity standardization, we normalized the intensities to have zero mean and standard deviation of 1.

We randomly split the 95 patients to create our train, validation, and test sets with 66, 9, and 20 patients respectively. After horizontal flipping based data augmentation, the train and validation sets had 700 and 106 slices respectively. The test set included 139 slices, 24 cancerous lesions, 1.12M pixels in the prostate with $9\%$ cancer pixels. We performed MRI scale standardization on the train set, and used the learned histograms to standardize the validation and test sets. We followed a similar strategy for MRI intensity normalization.

\subsection{Learning correlated features}\label{sec:corrnet}
\textbf{Feature extraction:} We extracted features from the T2W, ADC, and histopathology images by passing them through the first two convolutional layers of a pre-trained VGG-16 architecture \cite{simonyan2014very}. Thus, each $224 \times 224$ image yielded a $224 \times 224 \times 64$ representation, generating 64 features per pixel. We sampled pixels from within the prostate, and concatenated the T2W and ADC features to form the MRI representation per pixel. Thus, for each pixel, we had the MRI representation $\mathcal{R}_i \in \mathbb{R}^{128}$ and the histopathology representation $\mathcal{P}_i \in \mathbb{R}^{64}$.\\
\textbf{Common Representation learning:} We trained a Correlational Neural Network architecture (CorrNet) \cite{chandar2016correlational} to learn common representations from MRI and histopathology features per pixel. Given N pixels, each pixel input ${Z}_i$ to the CorrNet model had two views: the MRI feature representation for pixel $i$, $\mathcal{R}_i$, and the histopathology feature representation for pixel $i$, $\mathcal{P}_i$. We used a fully-connected CorrNet model with a single hidden layer, where the hidden layer $H(Z_i) \in \mathbb {R}^k$  was computed as: 
\begin{dmath}
H(Z_i) = W \mathcal{R}_i + V \mathcal{P}_i  + b 
\end{dmath}
where $W \in \mathbb{R}^{k \times 128}$, $V \in \mathbb{R}^{k \times 64}$ and $b \in \mathbb{R}^{k \times 1}$. The reconstructed output $Z_i'$ was computed from the hidden layer as:
\begin{dmath}
Z_i'=[W'H(Z_i),V'H(Z_i)]+b'
\end{dmath}
where $W' \in \mathbb{R}^{128 \times k}$, $V' \in \mathbb{R}^{64 \times k}$ and $b' \in \mathbb{R}^{(128+64) \times 1}$. In contrast to the original CorrNet model, we did not use any non-linear activation function. We learned the parameters $\theta = \{W,V, W',V',b,b'\}$ of the system by minimizing the following objective function, as detailed in \cite{chandar2016correlational}:
\begin{dmath} \label{eq:corrnet_loss}
J(\theta) = \sum_{i=1}^N [L(Z_i,H(Z_i))
+L(Z_i, H(R_i))+L(Z_i,H(P_i))
-\lambda corr(H(R_i),H(P_i)) ]
\end{dmath}

\begin{dmath}
corr(H(R_i),H(P_i))\\
= \frac{\sum_{i=1}^N [(H(R_i)-\overline{H(R)})(H(P_i)-\overline{H(P)]}}{\sqrt{\sum_{i=1}^N(H(R_i)-\overline{H(R)})^2 \sum_{i=1}^N(H(P_i)-\overline{H(P)})^2}}
\end{dmath}
where $L$ is the reconstruction error, $\lambda$  is the scaling parameter to determine the relative weight of the correlation error with respect to the reconstruction errors, $\overline{H(R)}$ is the mean hidden representation
of the $1^{st}$ view and $\overline{H(P)}$  is the mean hidden representation
of the $2^{nd}$ view. Thus, the CorrNet model (i) minimizes the self and cross reconstruction errors, and (ii) maximizes the correlation between the hidden representations of the two views. Training CorrNet using pixel representations from within the prostate gave ample training samples to optimize the model, and to learn differences between cancer and non-cancer pixels.

\begin{figure}[!htbp]
\centering
\includegraphics[width = 0.8\linewidth]{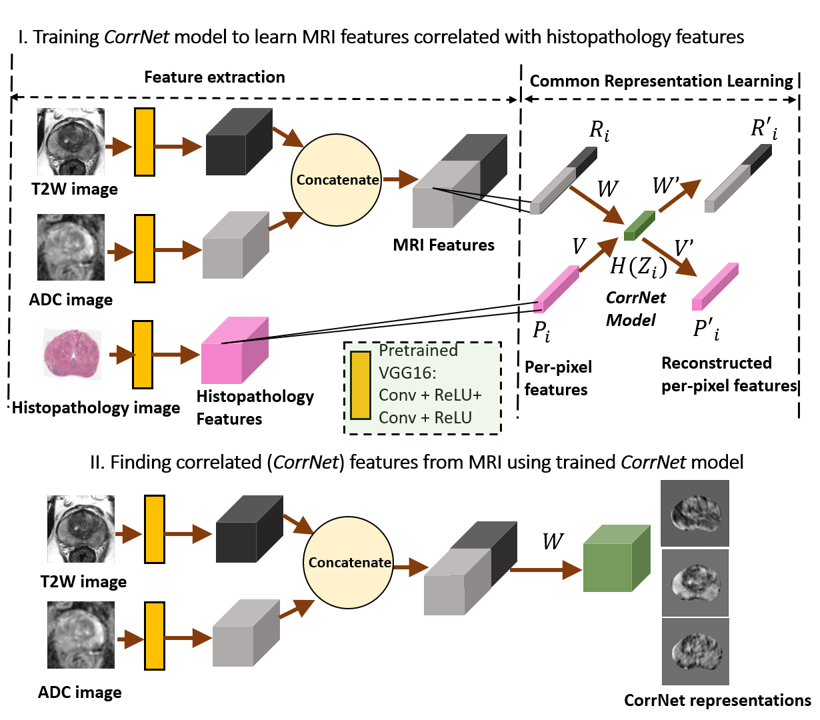}
\caption{Learning correlated representations from spatially aligned MRI and histopathology images, and then constructing the correlated (\textit{CorrNet}) representations from MRI alone using learned weights.}
\label{fig:corrnet}
\end{figure}

After the CorrNet model was trained, we used the learned weights $W,b$ to project the MRI feature representations $\mathcal{R}_i$ onto the $k$ dimensional hidden space to form \textit{CorrNet representations} of the input MRI. The \textit{CorrNet representations} 
are correlated with the corresponding histopathology features, and once trained, can be constructed even in the absence of histopathology images. Figure \ref{fig:corrnet} shows the pipeline for learning common representations.\\
\begin{figure}[!htbp]
\centering
\includegraphics[trim=120 70 100 70,clip, width = 0.18\linewidth]{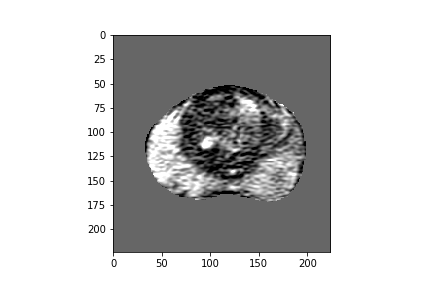}
\includegraphics[trim = 120 70 100 70,clip, width = 0.18\linewidth]{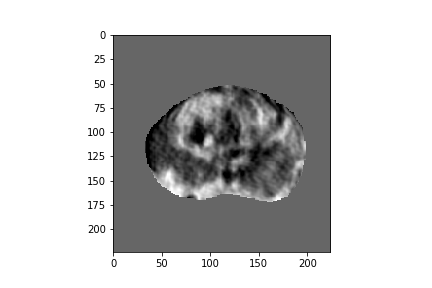}
\includegraphics[trim = 120 70 100 70,clip, width = 0.18\linewidth]{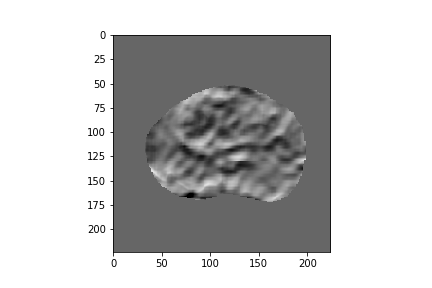}
\includegraphics[trim = 120 70 100 70,clip, width = 0.18\linewidth]{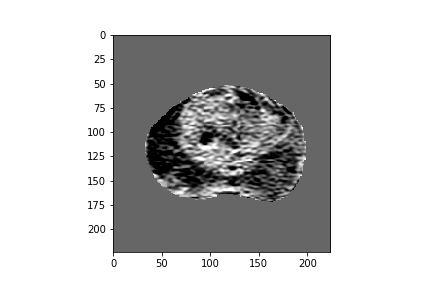}
\includegraphics[trim = 120 70 100 70,clip, width = 0.18\linewidth]{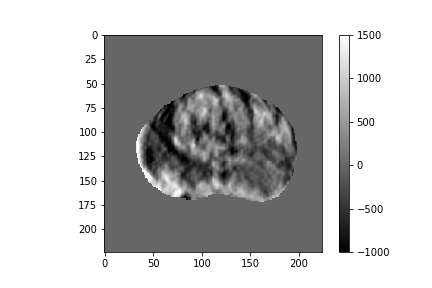}
\includegraphics[width = 0.033\linewidth]{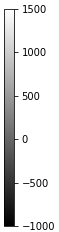}
\caption{Five-dimensional \textit{CorrNet representations} for one example MRI slice}
\label{fig:corr_rep}
\end{figure}
\begin{figure}[!htbp]
\centering
\includegraphics[trim = 0 5 0 0, clip, width = 0.8\linewidth]{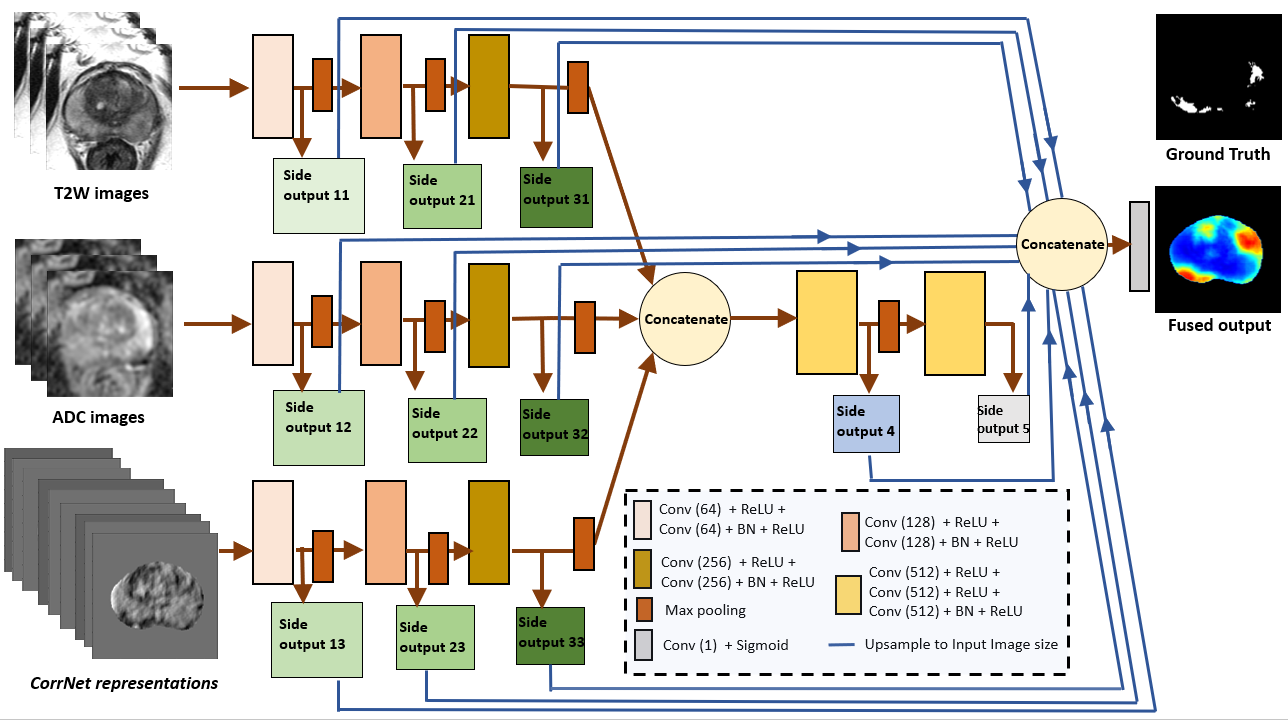}
\caption{HED-branch-3 model for predicting cancer probability maps.}
\label{fig:hedbranch}
\end{figure}
\textbf{Training:} From the 66 patients in the training cohort, we sampled all the cancer pixels from within the prostate, and randomly sampled an equal number of non-cancer pixels, also from within the prostate, thereby generating a training set of $\approx 1.2M$ pixels, with equal number of cancer and non-cancer pixels. This ensured that we train the CorrNet with a balanced dataset of two classes. We used $\lambda = 2$ to weigh the cross-correlation error higher than the reconstruction errors. We chose a squared error loss $L$ for the reconstruction errors. We trained the CorrNet model with varying hidden layer dimensions, namely: $k \in \{ 1, 3, 5, 15, 30\}$. For each $k$, we used a learning rate $\eta = 10^{-5}$, and 300 training epochs. Figure \ref{fig:corr_rep} shows \textit{CorrNet} representations of an example MRI slice, with $k=5$.
\subsection{Prediction of prostate cancer extent}\label{sec:pred}
We modified the Holistically Nested Edge Detection (HED) architecture \cite{HED}  to predict cancer probability maps for the entire prostate. We considered two modified versions of HED: (1) HED-3, and (2) HED-branch-3. The HED-3 model evaluates how well \textit{CorrNet representations} alone perform in predicting cancer, while the HED-branch-3 model evaluates how well \textit{CorrNet representations} combined with T2W and ADC images perform in predicting cancer. We represent our model using correlated feature learning and HED-3 as CorrSigNet($k$), and our model with correlated feature learning and HED-branch-3 as CorrSigNet(T2W, ADC, $k$), where $k$ is the \textit{CorrNet} feature dimension. For example, CorrSigNet($5$) uses only 5 correlated features for prediction, whereas CorrSigNet (T2W, ADC, $5$) uses
the normalized T2W and ADC intensities in addition to 5 correlated features for prediction. We chose a prediction model similar to the HED architecture because it is known to learn and combine multi-scale and multi-level features, and has been successfully applied to anatomy segmentations from CT scans \cite{harrison2017progressive, roth2016spatial, nogues2016automatic}, and to prostate cancer prediction \cite{sumathipala2018prostate}. 

In HED-3, we input three adjacent \textit{CorrNet slice representations} of the prostate and output predictions for only the central slice. This ensured that the 2D-HED model learned the 3D volumetric continuity from MRI/ histopathology/ correlated features. This also helped in reducing false positive rates.  

In HED-branch-3 (shown in Figure \ref{fig:hedbranch}), we combined the \textit{CorrNet slice representations} together with the normalized T2W and ADC images as inputs to the model. Similar to HED-3, we considered three adjacent slices for each input sequence (T2W, ADC, \textit{CorrNet representations}), and predicted cancer probability maps for the central slice only. However, in HED-branch-3 model, we processed each input sequence independently using the first three blocks, concatenated the three outputs from the three independent blocks, and processed the concatenated output using the next 2 blocks. 
Since the input sequences are processed independently in the first three blocks, we had a total of 11 side outputs, which were fused together using a Conv-1D layer to form the weighted fused output. We computed balanced cross-entropy losses for each of the 12 outputs (11 side outputs and 1 fused output) while training the architecture, but computed evaluation metrics only on the fused output. We used $3 \times 3$ kernels for all convolution layers except the last Conv-1D layer. The number of filters in each layer is stated in the legend in Figure \ref{fig:hedbranch}. For both the HED-3 and HED-branch-3 models, we added Batch Normalization in each block, before ReLU activation, as opposed to the HED model used by \cite{sumathipala2018prostate} which used Batch Normalization in each layer. No post-processing steps were
performed on the prediction maps.\\
\textbf{Training:} We trained both models using an Adam optimizer with an initial learning rate $\eta = 10^{-3}$, weight decay $\alpha = 0.1$, epochs = 200 and early stopping.
\section{Experimental Results}
\begin{figure}[!htbp]
\centering
\includegraphics[trim=70 90 60 80,clip, width = 0.25\linewidth]{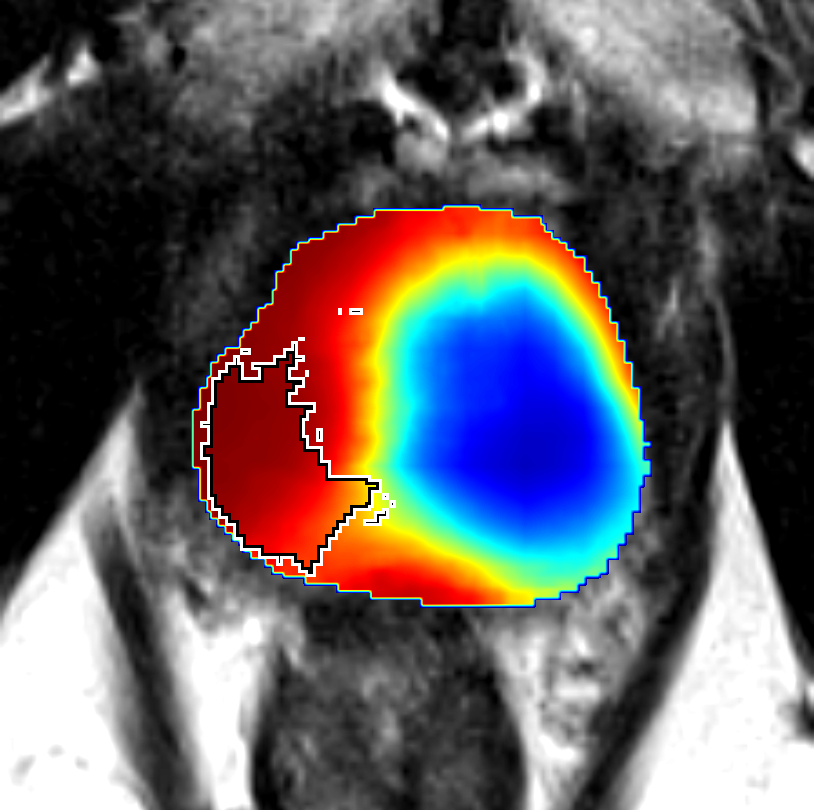}
\includegraphics[trim=70 90 60 80,clip, width = 0.25\linewidth]{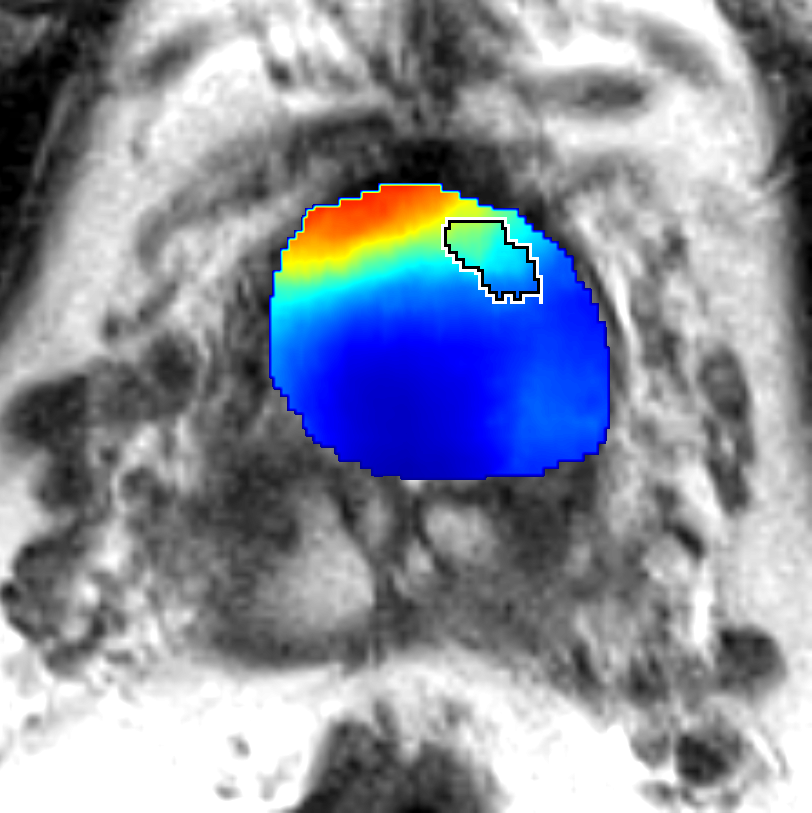}
\includegraphics[trim=70 90 60 80,clip, width = 0.25\linewidth]{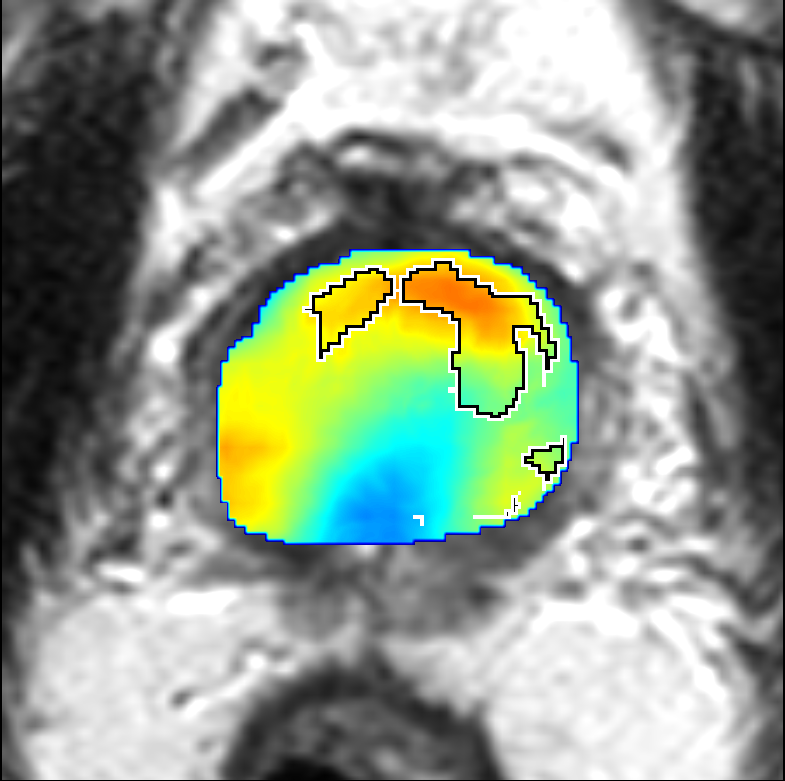}
\includegraphics[trim=70 90 60 80,clip, width = 0.25\linewidth]{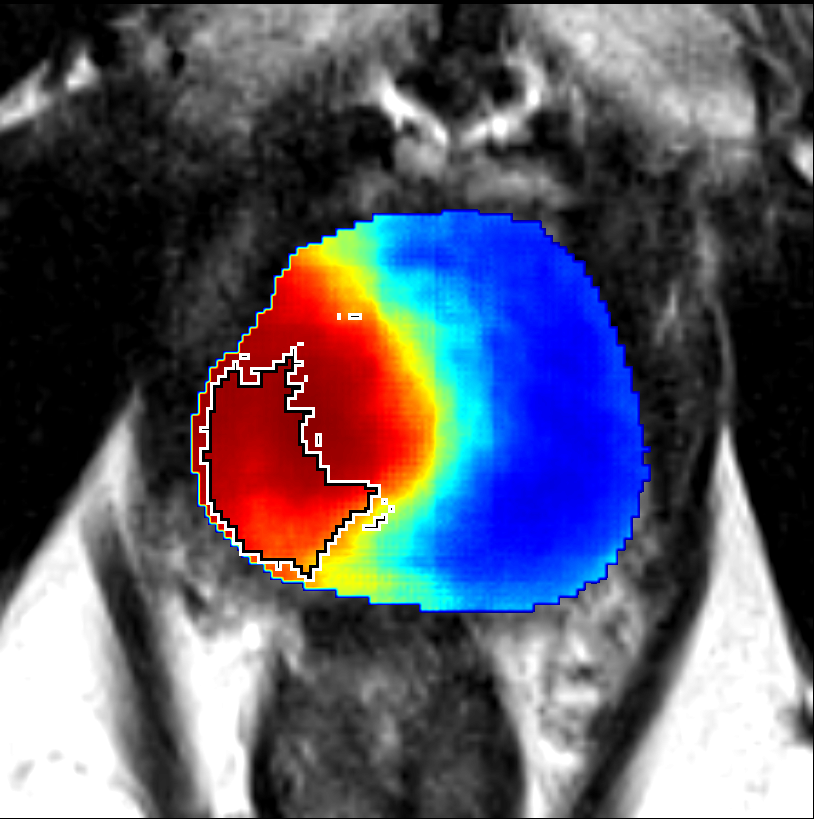}
\includegraphics[trim=70 90 60 80,clip, width = 0.25\linewidth]{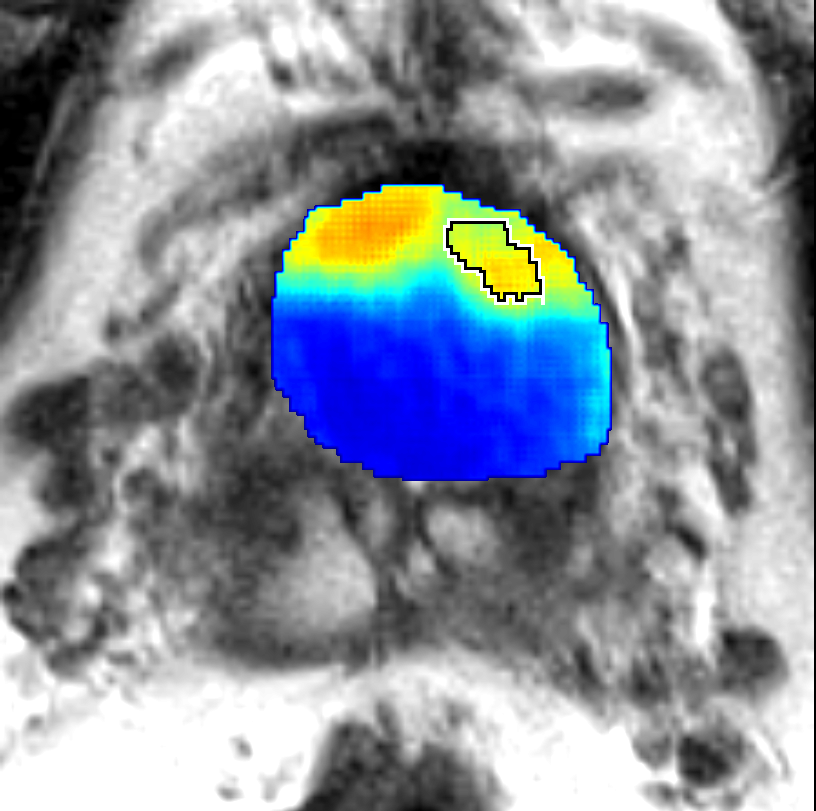}
\includegraphics[trim=70 90 60 80,clip, width = 0.25\linewidth]{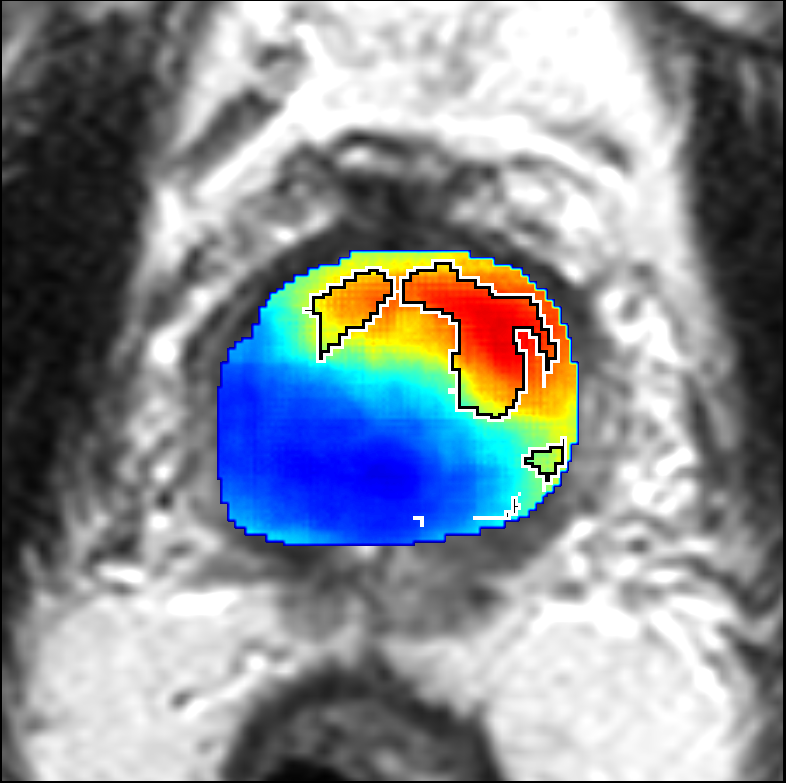}
\caption{(Top) Prediction results using the current state-of-the-art method \cite{sumathipala2018prostate}. (Bottom) Prediction results from our model CorrSigNet(T2W, ADC, $5$).}
\label{fig:res2}
\end{figure}

\textbf{Quantitative Evaluation:} We quantitatively evaluated our models on a per-pixel and a per-lesion basis, with ground truth labels derived from pathologist cancer annotations on registered histopathology images. For a direct comparison, we reproduced the current state-of-the-art model \cite{sumathipala2018prostate} to the best of our understanding, and computed both pixel-level and lesion-level evaluation metrics of this model on our test data (20 patients, 139 slices, 24 cancerous lesions, 1.12M pixels in the prostate). It may be noted that the AUC numbers reported in \cite{sumathipala2018prostate} are computed on a lesion level, and not on a pixel-level. Our pixel-level metrics including all pixels within the prostate provide a more rigorous evaluation.\\
\begin{table}[!htbp]
\centering
\caption{Pixel-level quantitative evaluation of CorrSigNet models}\label{tab:results}
\begin{tabular}{|c|c|c|c|}
\hline
\textbf{Model} &  \textbf{Sensitivity}  & \textbf{Specificity} & \textbf{AUC}\\
\hline
HED \cite{sumathipala2018prostate} (current state-of-the-art) & 0.75 & 0.74 & 0.80\\
CorrSigNet(1) &  0.72 & 0.78 & 0.81\\
CorrSigNet(3) & 0.82 & 0.71 & \textbf{0.86}\\
CorrSigNet(5) & 0.77 & 0.77 & \textbf{0.86}\\
CorrSigNet(15) & 0.76  & 0.78 & 0.84\\
CorrSigNet(30) & 0.75 &  0.81 & 0.85\\
CorrSigNet(T2W, ADC, 1) & 0.73 & 0.79 & 0.83\\
CorrSigNet(T2W, ADC, 3)  & 0.70 & \textbf{0.85} & \textbf{0.86}\\
\textit{CorrSigNet(T2W, ADC, 5)} & \textit{0.81} & \textit{0.72} & \textit{\textbf{0.86}}\\
CorrSigNet(T2W, ADC, 15) & 0.78 & 0.78  & \textbf{0.86}\\
CorrSigNet(T2W, ADC, 30)  & \textbf{0.83} & 0.71 & \textbf{0.86}\\
\hline
\end{tabular}
\end{table}
\textit{Pixel level analysis:} We tested the performance of the CorrSigNet models with different inputs and varying \textit{CorrNet} feature dimension $k$ using the following pixel-level evaluation metrics (computed using 1.12M pixels in the prostate): sensitivity, specificity, and AUC of the ROC curve, with a probability threshold of 0.5. We note from \tablename~\ref{tab:results} that CorrSigNet performs better than \cite{sumathipala2018prostate}, with consistently higher AUC numbers in pixel-level analysis. The sensitivity and specificity numbers vary within the models. Our tests showed that at least 3 CorrNet features were necessary for improved performance over MRI alone. We chose CorrSigNet (T2W, ADC, $5$) as the optimum model, because it had high sensitivity, specificity and AUC, with an optimum number of parameters. Between false positives and false negatives, we note that a false negative is more detrimental than a false positive in the task of cancer prediction.\\
\textit{Lesion level analysis:} We performed lesion-level analysis using the evaluation method detailed in \cite{sumathipala2018prostate} and found that CorrSigNet(T2W, ADC, $5$) achieved a per-lesion AUC of $0.96\pm0.07$ compared to a per-lesion AUC of $0.92\pm0.09$ by \cite{sumathipala2018prostate} on the same test set.\\
\begin{figure}[!htbp]
\centering
\begin{subfigure}[b]{.25\linewidth}
  \centering
   \includegraphics[trim=50 90 40 80,clip, width = \linewidth]{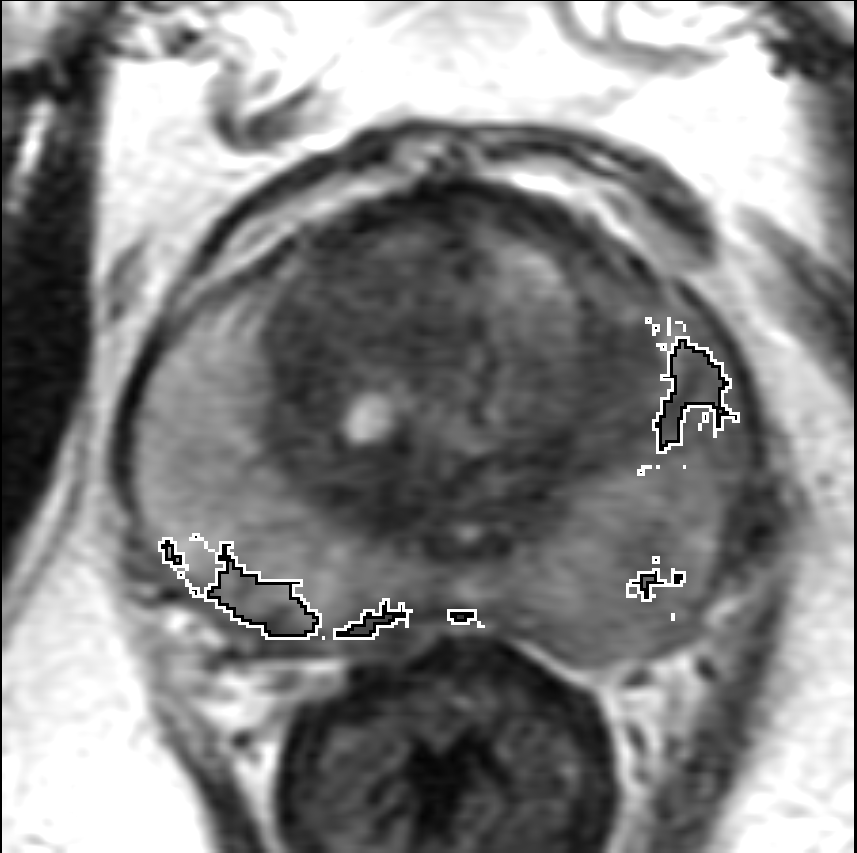}
  \caption{}
  \end{subfigure}
\begin{subfigure}[b]{.25\linewidth}
  \centering
 \includegraphics[trim=50 90 40 80,clip, width = \linewidth]{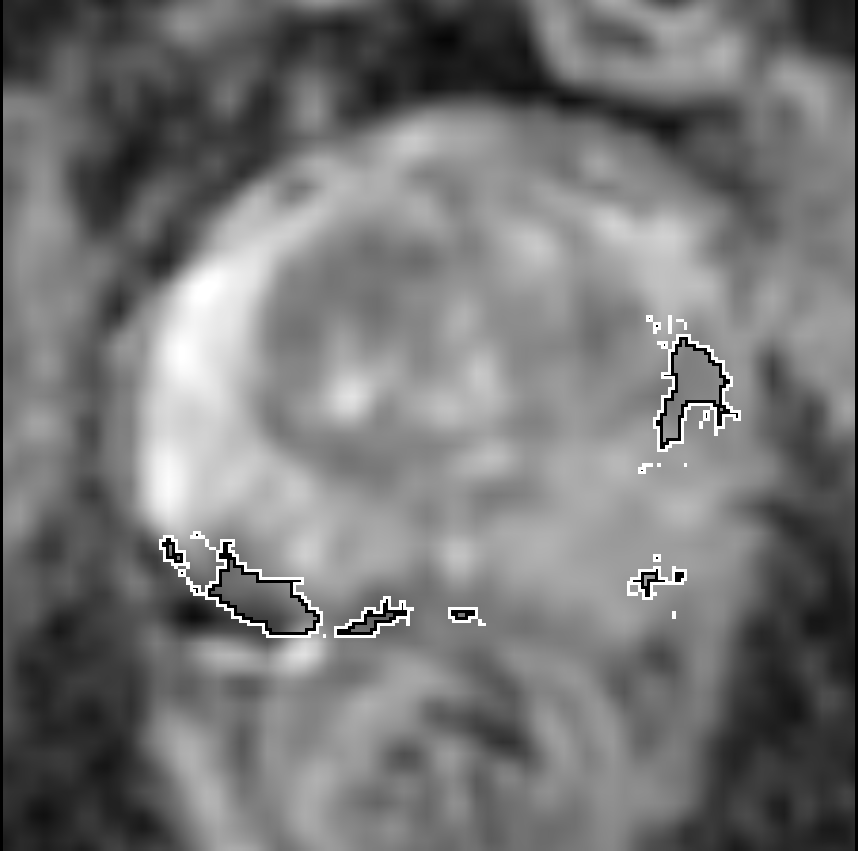}
  \caption{}
 \end{subfigure}
\begin{subfigure}[b]{.3\linewidth}
  \centering
 \includegraphics[trim=50 90 40 80,clip, width = 0.82\linewidth]{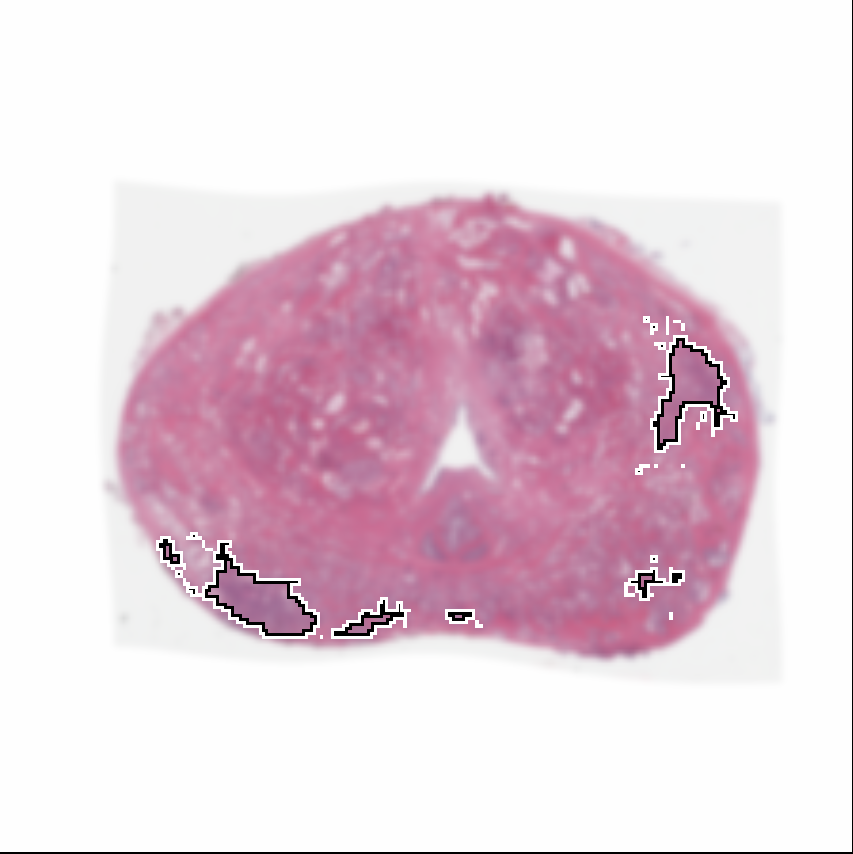}
 \includegraphics[width = 0.13\linewidth]{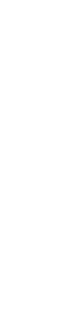}
  \caption{}
  \end{subfigure}
\begin{subfigure}[b]{.25\linewidth}
  \centering
 \includegraphics[trim=50 90 40 80,clip, width = \linewidth]{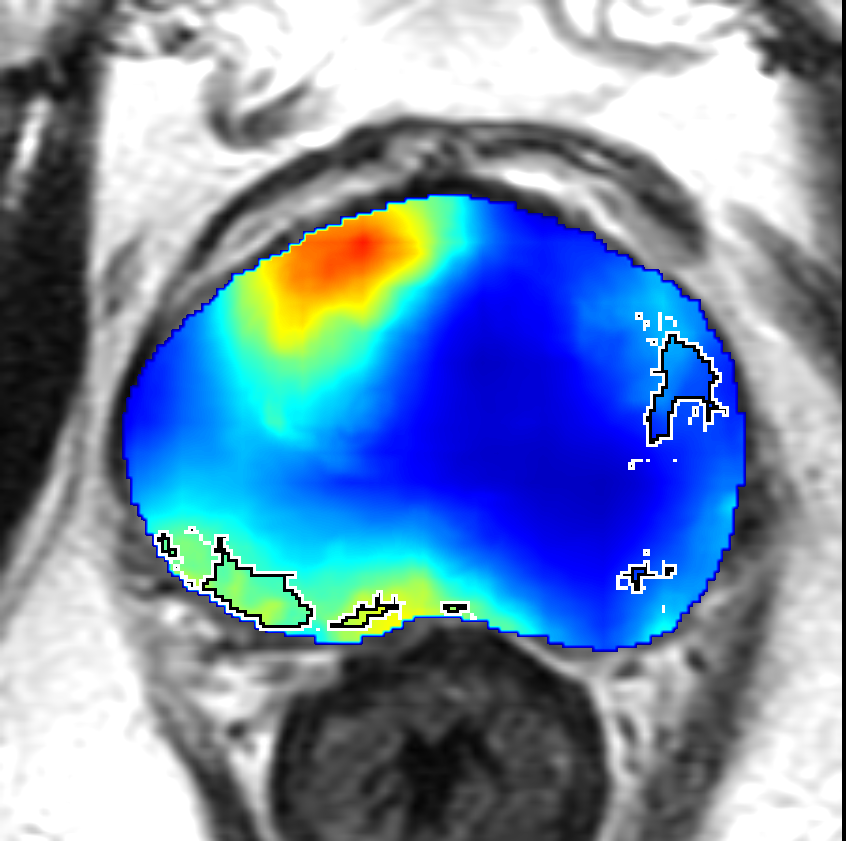}
  \caption{}
  \end{subfigure}
 \begin{subfigure}[b]{.25\linewidth}
  \centering
  \includegraphics[trim=50 90 40 80,clip, width = \linewidth]{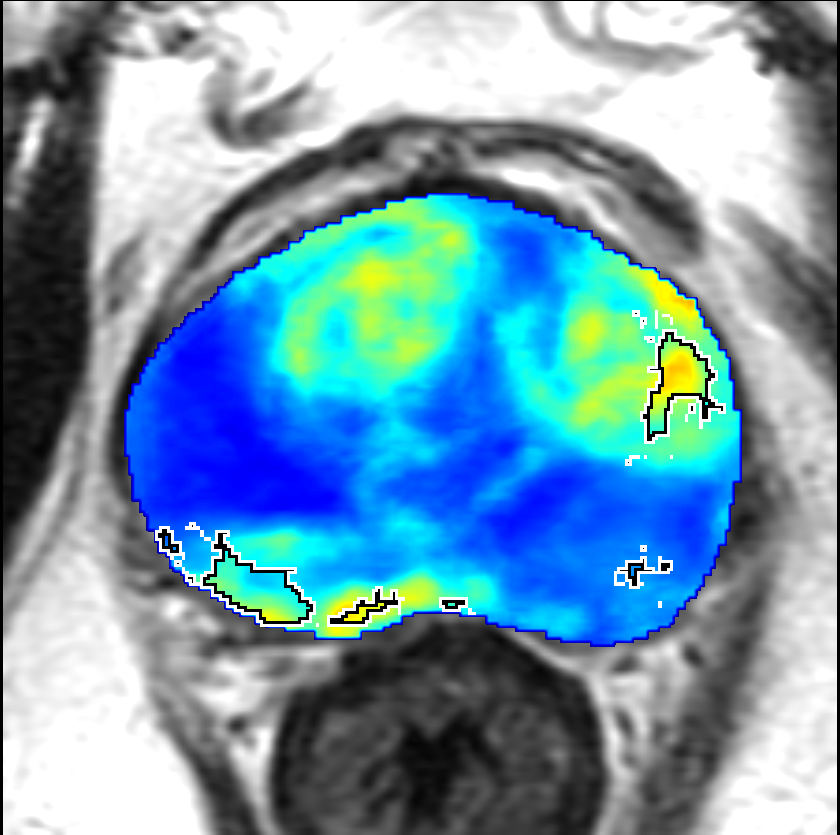}
  \caption{}
  \end{subfigure}
 \begin{subfigure}[b]{.3\linewidth}
  \centering
 \includegraphics[trim=50 90 40 80,clip, width = 0.82\linewidth]{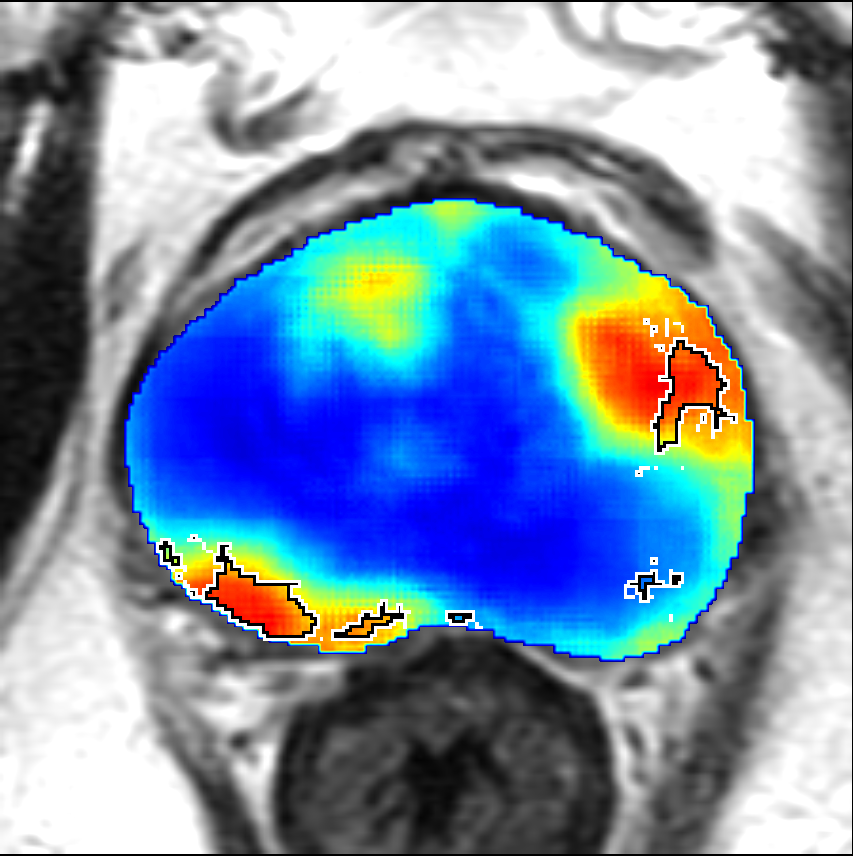}
 \includegraphics[width = 0.13\linewidth]{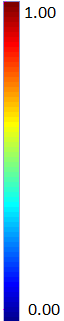}
   \caption{}
  \end{subfigure} 
\caption{Spatially aligned (a) T2W, (b) ADC, and (c) histopathology images. Data obtained and processed as detailed in \ref{sec:preprocess}. Prediction results using (d) the current state-of-the-art method \cite{sumathipala2018prostate}, (e) our model CorrSigNet($5$), and (f) our model CorrSigNet(T2W, ADC, $5$).}
\label{fig:res1}
\end{figure}
\textbf{Qualitative Evaluation:} Figure \ref{fig:res1} shows the same slice as in Figure \ref{fig:corr_rep} with aligned T2W, ADC, and histopathology images, and prediction results using current state-of-the-art method \cite{sumathipala2018prostate}, our CorrSigNet($5$) and CorrSigNet(T2W, ADC, $5$) models. It may be noted that \cite{sumathipala2018prostate} fails to detect the cancerous regions on the left and right of the images, while the \textit{CorrNet} representations alone can identify the cancer regions, and when combined with T2W and ADC images, they predict the cancer regions with high probability. It may also be noted that CorrSigNet(T2W, ADC, $5$) shows fewer false positives than \cite{sumathipala2018prostate}. This example shows the strength of learning correlated MRI signatures in identifying subtle, and sometimes MRI-invisible cancers. Figure \ref{fig:res2} shows more example slices from different patients, comparing the state-of-the-art approach \cite{sumathipala2018prostate} and our prediction results with CorrSigNet(T2W, ADC, $5$). We note that our model with correlated features (1) can identify subtle and smaller cancer regions, (2) have better overlap with ground truth cancer labels, and (3) have fewer false positives. 

\section{Conclusion}
In this paper, we presented a novel method to learn correlated signatures of cancer from spatially aligned MRI and histopathology images of prostatectomy surgical specimens, and then use these learned correlated signatures in predicting prostate cancer extent from MRI. Quantitatively, our method improved performance of automated prostate cancer localization (per-pixel AUC of 0.86, per-lesion AUC of $0.96 \pm 0.07$), as compared to the current state-of-the-art method \cite{sumathipala2018prostate} (per-pixel AUC 0.80, per-lesion $0.92 \pm 0.09$). Qualitatively, we found that correlated features could capture subtle cancerous regions and sometimes MRI-invisible cancers, had better overlap with ground truth labels, and fewer false positives. Correlated features have the capability of capturing tumor biology information from histopathology images in an unprecedented way, and these features, once learned, can be extracted in patients without histopathology images. In future work, we intend to conduct experiments with augmented datasets and in a cross-validation framework to boost the performance of our models.
\bibliographystyle{splncs04_sort}
\bibliography{refs}

\end{document}